\documentclass[runningheads]{llncs}
\usepackage[T1]{fontenc}
\usepackage{graphicx}
\usepackage{amsmath}
\usepackage{amssymb}
\usepackage{booktabs}
\usepackage{siunitx}

\usepackage{orcidlink}
\usepackage{hyperref}
\hypersetup{
pdfborder={0 0 0}  
}
\definecolor{scienceRed}{HTML}{D55E00} 
\definecolor{scienceSkyBlue}{HTML}{56B4E9} 
\definecolor{scienceBlue}{HTML}{0072B2} 
\definecolor{scienceGreen}{HTML}{009E73} 
\definecolor{scienceOrange}{HTML}{E69F00} 
\definecolor{sciencePurpur}{HTML}{CC79A7} 

\definecolor{atlasRed}{HTML}{BA4A32}
\definecolor{atlasRedMuted}{HTML}{F2DCD7}
\definecolor{atlasGreen}{HTML}{417501}
\definecolor{atlasGreenMuted}{HTML}{CFDCBF}
\definecolor{atlasBlue}{HTML}{074D6D}
\definecolor{atlasBlueMuted}{HTML}{C1D2DA}
\definecolor{atlasGrey}{HTML}{808080}
\definecolor{atlasGreyMuted}{HTML}{C5C5C5}
\definecolor{atlasGreyDark}{HTML}{222A35}

\usepackage{graphicx}
\usepackage{tikz}
\usetikzlibrary{3d, arrows.meta, positioning, calc, decorations.markings, angles, quotes, shapes.geometric, shapes.arrows, intersections, overlay-beamer-styles, tikzmark}
\usepackage{pgfplots}
\pgfplotsset{compat=newest,/pgf/number format/.cd,1000 sep={}}
\pgfplotsset{colormap={modrainbow}{rgb255(0cm)=(0,0,255); rgb255(1cm)=(0,255,255); rgb255(2cm)=(0,255,0);
rgb255(3cm)=(255,255,0); rgb255(4cm)=(255,0,0)}}
\usepackage{wasysym,amssymb}

\newif\ifarxiv
\arxivtrue 

\ifarxiv
  
\else
  \usetikzlibrary{external}
\fi

\begin{document}
\title{Accurate PIC Simulation of Ion Optics using p-Adaptation on Tetrahedral Meshes}
\titlerunning{Accurate PIC Simulation of Ion Optics using p-Adaptation on Tetr. Meshes}

\author{Kim-Sophie Ellenberger\inst{1}\orcidlink{0009-0006-1243-2822} \and
Tobias Ott\inst{1}\orcidlink{0009-0000-4038-0085} \and
Lasse Voigt\inst{1}\orcidlink{0009-0004-3852-3680} \and\\
Stefanos Fasoulas\inst{1}\orcidlink{0000-0002-4168-2410} \and
Marcel Pfeiffer\inst{1}\orcidlink{0000-0002-4087-1789}}
\authorrunning{K.-S.~Ellenberger, T.~Ott, L.~Voigt, S.~Fasoulas, and M.~Pfeiffer}
%
\institute{Institute of Space Systems (IRS), University of Stuttgart, \\Pfaffenwaldring 29, 70569 Stuttgart\\
\email{ellenbergerk@irs.uni-stuttgart.de}}

\maketitle              
\begin{abstract}

Satellites operating in Very Low Earth Orbit (VLEO) experience significant atmospheric drag and material erosion caused by impinging atomic oxygen (AO). To reproduce VLEO-relevant AO impact conditions on a laboratory scale, a negative-ion beamline is being developed.
A key component of this beamline is an electrostatic deceleration unit, which reduces the kinetic energy of the O$^{-}$ ions from \SI{20}{\kilo\electronvolt} to \SI{5}{\electronvolt} using a linearly decreasing electrical potential. Accurately resolving the resulting ion optics requires a numerical model that can handle complex, curved geometries.
In this work, we present a full three-dimensional Particle-In-Cell (PIC) simulation of the deceleration unit performed with the Plasma Kinetic Code PICLas, using unstructured tetrahedral meshes discretized with a continuous Galerkin field solver and a leapfrog time integrator. To balance accuracy and computational cost, we additionally employ local polynomial order (p-)adaption of the field solution, increasing the polynomial degree only in regions of high potential or charge-density gradients. Results of the transverse velocity for fixed polynomial orders $p=1$ to $p=4$ are compared against a locally p-adapted simulation. While $p=1$ severely underresolves the transverse velocity spread of the beam, $p\geq2$ simulations converge to a consistent velocity distribution. The p-adapted simulation reproduces this converged solution while reducing the runtime by about 33\,\% compared to a uniform $p=4$ simulation, demonstrating that local p-adaption is an effective strategy for efficient, high-fidelity PIC simulation of ion optics on unstructured tetrahedral meshes.

\keywords{Very Low Earth Orbit (VLEO) \and Particle-In-Cell (PIC) \and Ion Optics \and p-Adaption \and Tetrahedral Meshes \and Continuous Galerkin (CG)}

\end{abstract}

\section{Introduction}

Satellites operating in Very Low Earth Orbit (VLEO) are exposed to a residual atmosphere that is dominated by atomic oxygen (AO). The resulting aerodynamic drag causes rapid orbital decay, while AO impingement additionally erodes and chemically alters exposed satellite surface materials \cite{crisp2020benefits}. Understanding and mitigating these effects is essential for enabling future VLEO satellite platforms.
As part of this effort, an Atomic Oxygen Ground Testing Facility is being set up at the Institute of Space Systems, University of Stuttgart, to reproduce VLEO-relevant AO impact conditions under controlled laboratory conditions \cite{fasoulas2025motivation,ellenberger2025numerical}.

\begin{figure}[h]
    \centering
    \resizebox{1.0\textwidth}{!}{
    \includegraphics{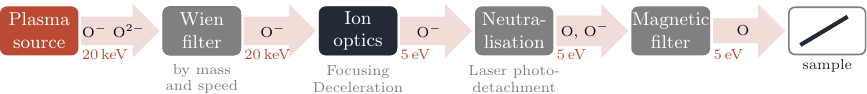}
    }
    \caption{Setup of the negative ion beamline to realize VLEO-relevant velocities. The numerical simulations in this paper focus on the deceleration unit from \SI{20}{\kilo\electronvolt} to \SI{5}{\kilo\electronvolt}.}
    \label{fig:setup_neg_ion_beamline}
\end{figure}

One idea of a ground testing facility is a negative-ion beamline that realizes VLEO-relevant AO impact energies. Figure\,\ref{fig:setup_neg_ion_beamline} shows the general setup and its consecutive steps to achieve VLEO conditions. A plasma source generates negative oxygen ions O$^{-}$ which are accelerated to \SI{20}{\kilo\electronvolt}. A Wien filter subsequently selects O$^{-}$ ions by mass and speed, removing all residual ions. The resulting O$^{-}$ beam is then decelerated by an ion-optical deceleration unit down to a VLEO-relevant energy of \SI{5}{\electronvolt}. The process of deceleration of the ion kinetic energy from \SI{20}{\kilo\electronvolt} to \SI{5}{\electronvolt} is realized with a static electric field with a linearly decreasing potential along the flight path and the numerical simulation thereof is the focus of this paper. The final \SI{5}{\electronvolt} energy is secured by two ion grids placed at the end of the deceleration stage. Downstream, laser photodetachment neutralizes a fraction of the O$^{-}$ beam into neutral atomic oxygen, and a magnetic filter removes the remaining charged particles before the neutral beam reaches the sample under test.

\vspace{-0.5cm}
\vspace{-0.5cm}

Accurate numerical simulations are essential for the design and optimization of experimental setups like ion optics. Particle-In-Cell (PIC) methods are widely used for this purpose, as they allow a self-consistent treatment of charged particle dynamics and electromagnetic fields. However, the complex geometrical structures typically present in ion optical setups pose significant challenges for mesh generation and numerical efficiency. In particular, the use of structured hexahedral meshes becomes problematic for geometries featuring intricate electrodes, curved surfaces, and narrow gaps. These limitations can lead to poor mesh quality, reduced numerical accuracy, or excessive computational cost due to mesh refinement algorithms.

An initial simulation attempt using the full, geometrically detailed cylindrical electrode stack revealed several practical obstacles: the series of more than 50 curved and narrow-gapped electrodes forms a geometrically complex domain, and a particle-in-cell (PIC) discretization on hexahedral meshes proved computationally expensive, with the direct field solver failing to converge for realistic mesh resolutions. To address this issue, the present work investigates the use of unstructured tetrahedral meshes as an alternative strategy for PIC simulations of ion optics with the open-source software PICLas \cite{fasoulas2019combining,munz2014coupled}, developed at the Institute of Space Systems, University of Stuttgart and the spin-off company boltzplatz. The Plasma Kinetic Code PICLas \cite{piclas_github} covers a multitude of methods with the possibility to combine them with each other, for example PIC and Direct Simulation Monte Carlo (DSMC).

\section{Numerical Method}

\begin{figure}
	\centering
	\includegraphics[width=\textwidth, trim=0pt 400pt 130pt 370pt,clip]{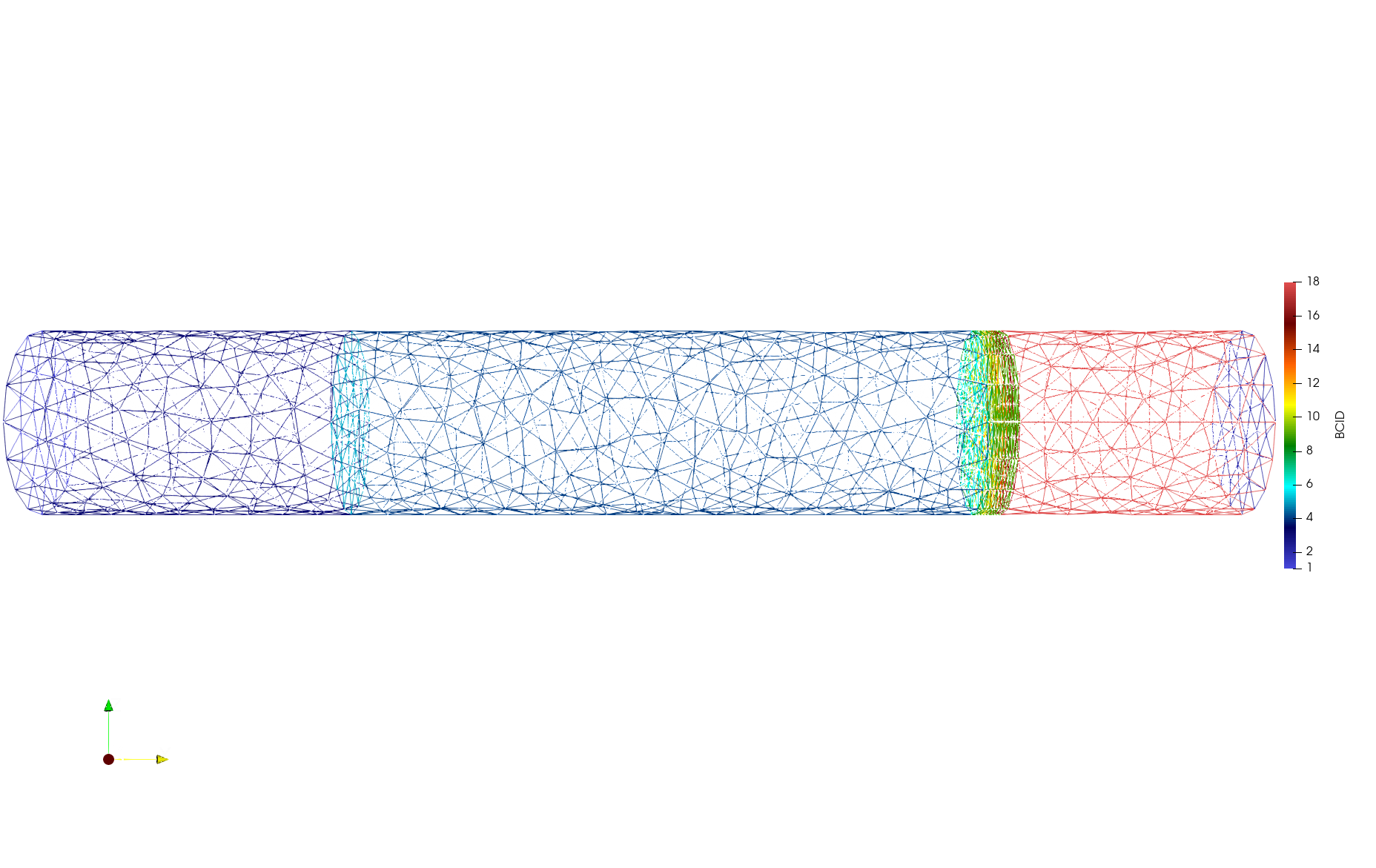}
    \vspace{-0.9cm}
    \caption{Simplified tetrahedral surface mesh used for all numerical simulations in this paper. The negative ion beam will be in the rotationally symmetry axis of the mesh. Deceleration unit with linearly decreasing potential colored in sky blue in the middle ending with the ion grid pair to realize the VLEO-relevant \SI{5}{\kilo\electronvolt} energy of the ions.}
    \label{fig:tet_mesh}
\end{figure}

As a first simplification, the beforehand intricate geometry was reduced to a set of coaxial cylindrical sections with a linearly decreasing potential instead of 50 stepwise decreasing constant potentials. In contrast to the previously used HDG solver \cite{pfeiffer2019particle}, which relies on a hybridizable discontinuous Galerkin discretization, the field solver used here employs a continuous Galerkin (CG) formulation on unstructured tetrahedral meshes, that guarantees energy conservation in the PIC scheme in the $\Delta t \rightarrow 0$ limit. The surface mesh structure is presented in Figure\,\ref{fig:tet_mesh}.

\vspace{-0.5cm}
\vspace{-0.5cm}

\subsection{Leapfrog Time Discretization}

The electrostatic PIC solver in PICLas advances particles and fields in time using a standard leapfrog scheme. Within one time step $\Delta t$, the electric field $\mathbf{E}^n$ is first updated by solving the Poisson equation; the field is then interpolated to the particle positions, $\mathbf{E}_p^n \leftarrow \mathbf{E}_i^n$. Particle velocities are advanced using a velocity (half-step leap) update,
\begin{equation}
\frac{\mathbf{v}^{n+\frac{1}{2}} - \mathbf{v}^{n-\frac{1}{2}}}{\Delta t} = \frac{q}{m}\mathbf{E}_p^n ,
\end{equation}
followed by a position update
\begin{equation}
\frac{\mathbf{x}^{n+1} - \mathbf{x}^n}{\Delta t} = \mathbf{v}^{n+\frac{1}{2}} .
\end{equation}
The updated particle positions and charges are then deposited on the mesh to obtain the charge density,
\begin{equation}
\rho_i^{n+1} \leftarrow q_p, \; \mathbf{x}_p^{n+1} ,
\end{equation}
which closes the loop by feeding into the next Poisson solve,
\begin{equation}
\nabla^2 \varphi = -\frac{\rho}{\varepsilon_0} .
\end{equation}

\subsection{Continuous Galerkin Field Solver with p-Adaption}

\begin{figure}[h]
    \centering
    \resizebox{0.8\textwidth}{!}{
    \includegraphics{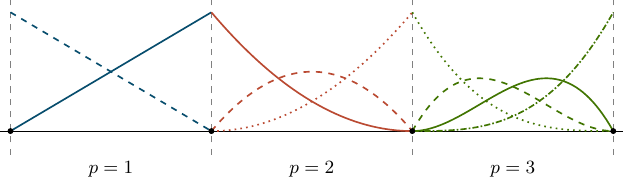}
    }
    \caption{Bernstein-B\'ezier basis functions per element; shared endpoint values at interfaces enforce continuity across the $p$-adapted interfaces.}
    \label{fig:bernstein_bezier}
\end{figure}

The electrostatic potential $\varphi$ is obtained from the Poisson equation
\begin{equation}
\nabla^2 \varphi = -\frac{\rho}{\varepsilon_0} \; \text{in } \Omega, \qquad
\begin{cases}
\varphi = \varphi^D, & x \in \partial\Omega^D \\[2pt]
\dfrac{\partial \varphi}{\partial n} = \mathbf{n}\cdot\mathbf{E} = E^N, & x \in \partial\Omega^N
\end{cases}
\end{equation}
with Dirichlet boundary conditions $\varphi^D$ on the electrode surfaces and Neumann boundary conditions $E^N$ elsewhere. With the test function $\Phi$, the corresponding weak form reads
\begin{equation}
\int_{\Omega} \nabla\varphi \cdot \nabla\Phi \, dx = \int_{\Omega} \frac{\rho}{\varepsilon_0} \Phi \, dx - \int_{\partial\Omega^N} E^N \Phi \, dx ,
\end{equation}
which is discretized on the unstructured tetrahedral mesh using a Continuous Galerkin (CG) method with Bernstein-B\'ezier basis functions defined locally per element, shown in Figure\,\ref{fig:bernstein_bezier}. Continuity of the field solution across element interfaces is enforced through shared degrees of freedom at the element boundaries.

\vspace{-0.5cm}

A central feature of the field solver is the possibility of local p-adaption: instead of using a single, globally uniform polynomial order $p$, the polynomial degree is chosen individually for each mesh element. The polynomial order is increased in regions where the gradient of the potential $\varphi$ or of the charge density $\rho$ is large, typically in regions of high particle density along the flight path of the ions, while it is kept low in regions where the field varies smoothly, such as before the deceleration unit. This allows the local resolution of the field representation to be matched to the local solution complexity, reducing the total number of degrees of freedom compared to a uniformly high polynomial order, while retaining high accuracy where it is needed \cite{ott2026p}.

Combined with the CG field discretization, the leapfrog-based PIC scheme in PICLas exhibits energy conservation in the limit $\Delta t \rightarrow 0$, which is an important prerequisite for efficient long-time, full simulations of the ion beam of the negative ion beamline with all steps.

\section{Results}

The p-adaption approach was applied to a three-dimensional PIC simulation of the simplified deceleration unit geometry (coaxial cylindrical electrode sections with linearly decreasing potential), tracking O$^{-}$ ions from \SI{20}{\kilo\electronvolt} down to \SI{5}{\electronvolt} on an unstructured tetrahedral mesh. Beforehand, the transverse velocity distribution is compared to simulations with a uniform polynomial order $p$.

\subsection{Effect of Uniform Polynomial Order}

\begin{figure}[h]
    \centering
    \resizebox{0.8\textwidth}{!}{
     \includegraphics{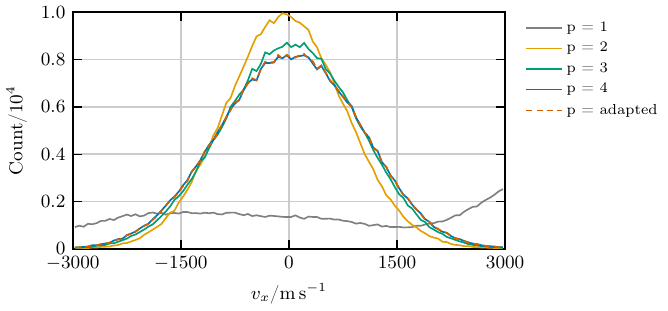}

    }
    \caption{Transverse velocity component sampled on a plane perpendicular to the beam axis for uniform polynomial orders of $p=1,2,3,4$ as well as an adapted polynomial degree between $p=1$ and $p=4$.}
    \label{fig:v_x_for_p}
\end{figure}

Simulations were first performed using a uniform polynomial order $p = 1, 2, 3$, and $4$ throughout the domain. For each case, the beam was sampled on a plane perpendicular to the beam axis near the end of the deceleration stage, and the distribution of the transverse velocity component $v_x$ was evaluated as shown in Figure\,\ref{fig:v_x_for_p}. The $p=1$ simulation produces a markedly broader and flatter velocity distribution than the higher-order cases, indicating that the linear field representation is too coarse to correctly resolve the focusing behavior of the ion optics at the low-energy end of the deceleration unit. In contrast, the $p=2$, $p=3$, and $p=4$ simulations yield transverse velocity distributions that are close to one another, indicating convergence of the field solution with increasing polynomial order. This improved accuracy, however, comes at a substantial computational cost: the measured runtimes increase from \SI{24}{\minute} for $p=1$ to \SI{2}{\hour}, \SI{5}{\hour}, and \SI{15}{\hour} for $p=2$, $p=3$, and $p=4$, respectively.

A qualitative comparison of the particle distributions for $p=1$ and $p=4$, colored by transverse velocity, illustrates the same effect directly in physical space: the $p=4$ simulation resolves a distinctly more focused, cone-shaped divergence of the beam downstream of the deceleration unit, consistent with the expected ion-optical behavior, whereas the $p=1$ simulation shows an unphysically broad, poorly structured particle spread.

\subsection{p-Adapted Simulation}

\begin{figure}
	\centering
    \includegraphics{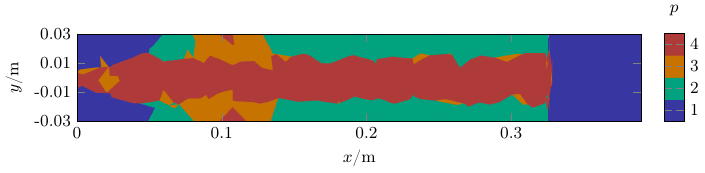}
    \caption{Slice along the beam axis of the locally adapted polynomial order. Along the flight path of the ions, a higher polynomial order is used ($p=4$), while in areas without space charge along the deceleration unit, a local polynomial order of $p=2$ proved to be sufficient.}
    \label{fig:p_adaption}
\end{figure}

Based on the local distribution of the potential gradient along the deceleration unit, a p-adapted simulation was performed in which the local polynomial order per element varies between 1 and 4: low polynomial orders are retained in the sections before and after the deceleration unit, while higher polynomial orders are assigned locally near the transition into the deceleration unit and along the flight path of the ions in the rotational symmetry axis of the deceleration unit, where space charge effects occur. Figure\,\ref{fig:p_adaption} displays the local polynomial order along the beam axis. The ragged edges are due to the slicing of the result, mostly not along the tetrahedral mesh edges.
\vspace{-0.5cm}
\vspace{-0.5cm}

The resulting transverse velocity distribution obtained with the p-adapted simulation closely matches the converged distributions obtained with uniform $p=4$, correctly reproducing the VLEO-relevant impact velocities of the decelerated ions and is also included in Figure~\ref{fig:v_x_for_p}. At the same time, the p-adapted simulation completes in approximately \SI{5}{\hour}, corresponding to a runtime reduction of about 33\,\% compared to the uniform $p=4$ simulation, while retaining comparable accuracy. Therefore, it can be confirmed that the locally reduced polynomial order does not compromise the overall solution quality while reducing the simulation runtime to a fraction of the simulation with uniform polynomial order.

\begin{figure}
\centering
\includegraphics{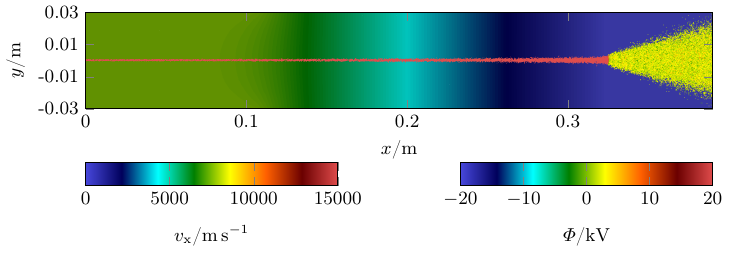}

    \caption{PIC simulation of the negative ion beamline with slice of potential in the beam axis. The particles are colored by their velocity in flight direction, showing VLEO-relevant velocities after exiting the deceleration unit, though space charge effects have led to an extensive broadening of the beam.}
    \label{fig:simulation_p_adapted}
\end{figure}

A close-up of the full simulation result with adapted polynomial order is presented in Figure\,\ref{fig:simulation_p_adapted}. A slice of the linearly decreasing electrical potential $\varphi$ is shown in the background along the beam axis.
The particles are colored by their velocity in flight direction $v_\text{y}$, which are continuously decelerated through the electric potential realizing VLEO-relevant velocities after exiting the deceleration unit. Increasing with decreasing velocities, space charge effects have led to an extensive broadening of the beam due to the charged particles interacting with each other.

\section{Summary and Outlook}

This work investigated a negative-ion beamline developed
for laboratory-scale reproduction of VLEO-relevant atomic oxygen impact conditions, focusing on the ion optics of its electrostatic deceleration unit. A three-dimensional PIC simulation of the deceleration unit was performed with the Plasma Kinetic Code PICLas using unstructured tetrahedral meshes, a continuous Galerkin field solver, and a leapfrog time integrator, avoiding the convergence and efficiency problems encountered with hexahedral discretizations. A comparison of simulations with uniform polynomial order $p=1$--$4$ showed that high polynomial orders are required to correctly resolve the transverse velocity distribution of the decelerated ion beam, at substantially increasing computational cost for higher uniform polynomial orders. Local p-adaption of the polynomial order, guided by the local gradient of the electrostatic potential and charge density, was shown to reproduce the accuracy of the converged high-order simulations while reducing the runtime by about 33\,\% relative to a uniform $p=4$ simulation.

Future work will focus on comparing the simulation results with experimental data from an upcoming measurement campaign on the negative-ion beamline, further optimization of the beamline and its individual components, and extending the simulation approach to further elements of the beamline.

\begin{credits}
\subsubsection{\ackname} This work is funded by the Deutsche Forschungsgemeinschaft (DFG, German Research Foundation) – Project-ID 516238647 – SFB 1667/1 (ATLAS - Advancing Technologies of Very Low-Altitude Satellites). The authors gratefully acknowledge the computing time provided on the high-performance computer HoreKa by the National High-Performance Computing Center at KIT (NHR@KIT). This center is jointly supported by the Federal Ministry of Research, Technology and Space and the Ministry of Science, Research and the Arts of Baden-Württemberg, as part of the National High-Performance Computing (NHR) joint funding program. HoreKa is partly funded by the German Research Foundation (DFG).

\subsubsection{\discintname}
The authors have no competing interests to declare that are relevant to the content of this article.
\end{credits}

%
%

\bibliographystyle{splncs04}%
\bibliography{bib}

\end{document}